\documentclass[twocolumn,epsfig,APS,floats,showpacs]{revtex4}
\usepackage{graphicx}
\usepackage{color}

\begin{document}

\title{Magnetic anisotropy in Cr$_2$GeC investigated by X-ray magnetic circular dichroism and \textit{ab initio} calculations}

\author{Martin Magnuson$^{1}$ and Maurizio Mattesini$^{2,3}$}

\affiliation{$^1$Department of Physics, Chemistry and Biology, IFM, Thin Film Physics Division, Link\"oping University, SE-58183 Link\"{o}ping, Sweden}

\affiliation{$^2$Department of Earth's Physics and Astrophysics, Complutense University of Madrid, Madrid, E-28040, Spain}

\affiliation{$^3$Instituto de Geociencias (CSIC-UCM), Facultad de CC. F\'isicas, E-28040 Madrid, Spain}

\date{\today}

\begin{abstract}
The magnetism in the inherently nanolaminated ternary MAX-phase Cr$_{2}$GeC is investigated by element-selective, polarization and temperature-dependent, soft X-ray absorption spectroscopy and X-ray magnetic circular dichroism. The measurements indicate an antiferro-magnetic Cr-Cr coupling along the $c$-axis of the hexagonal structure modulated by a ferromagnetic ordering in the nanolaminated $ab$-basal planes. The weak chromium magnetic moments are an order of magnitude stronger in the nanolaminated planes than along the vertical axis. Theoretically, a small but notable, non-spin-collinear component explains the existence of a non-perfect spin compensation along the $c$-axis. As shown in this work, this spin distortion generates an overall residual spin moment inside the unit cell resembling that of a ferri-magnet. Due to the different competing magnetic interactions, electron correlations and temperature effects both need to be considered to achieve a correct theoretical description of the Cr$_{2}$GeC magnetic properties.
\end{abstract}


\maketitle

\section{Introduction}

Ternary nanolaminated carbides and nitrides, known as $M_{n+1}AX_{n}$-phases, is the subject of intense research \cite{Barsoum1,Wang,Eklund1,Review}. Three related crystal structures are classified by stoichiometry as 211 ($n$=1), 312 ($n$=2) and 413 ($n$=3) phases, where the letter M denotes an early transition metal, A is an element in the groups III-V and X is carbon or nitrogen. The $M_{n+1}AX_{n}$-phases exhibit a technologically important combination of metallic and ceramic properties \cite{Barsoum_book}, that are related to the internal nanolaminated crystal structure, the choice of the three constituent elements, as well as the electronic structure and the chemical bonding between the intercalated atomic layers.

Macroscopic magnetism in MAX-phases was initially predicted by Luo \textit{et al.} in hypothetical Fe-containing 211 MAX-phases \cite{Ahuja}. Later, more accurate calculations including all competing phases have shown that this phase is thermodynamically unstable \cite{Dahlqvist}. Magnetic macroscopic response has also been observed in Cr$_{2-x}$Mn$_{x}$GaC phases \cite{Lin} that could be related to strong magnetic perovskite impurities. Recently, a magnetic macroscopic response has been observed in Mn-doped phase-pure (Cr,Mn)$_{2}$GeC at room temperature \cite{Ingason}. 

Although pure Cr$_{2}$GeC is macroscopically non-magnetic at room temperature \cite{Magnuson0}, the Cr contribution in Cr$_{2}$GeC has an anti-ferromagnetic (AFM), ferromagnetic (FM), ferrimagnetic (FiM), paramagnetic (PM) or non-magnetic (NM) order that depend on the temperature. Since the magnetic ordering could affect other properties of the MAX-phases, such as the thermal expansion and the bulk modulus, it is important to know the atomic magnetic exchange interactions in detail as non-magnetic calculations yield insufficient agreement with experiment \cite{Magnuson0,Grossi}. Experimentally obtained data of element-specific magnetic coupling are therefore important \cite{Lue,AA,Schneider,Durr,Stohr1,Stohr2}.
Moreover, the nature of the correlation effects of the localized Cr $3d$ states make the magnetic coupling theoretically complicated \cite{Ramzan1,Ramzan2,Maurizio,Jaouen}.

Previous investigations of Cr$_{2}$GeC include several theoretical studies, where the magnetic coupling in the electronic structure has been a controversial issue \cite{Zhou,Maurizio,Ramzan1,Ramzan2,Jaouen}. Using standard Density Functional Theory (DFT) within the PBE scheme, Zhou \textit{et al.} \cite{Zhou} found that the ground state at 0 K of Cr$_{2}$GeC is antiferromagnetic (AFM) while the ferromagnetic configuration is a metastable state. In the AFM case, a significant band spin-split of $\sim$ 2 eV was also predicted at the Fermi level ($E_{F}$) \cite{Zhou}. 

However, the Local Density Approximation (LDA) yields very close (degenerate) energy difference between FM and AFM ordering in Cr$_{2}$GeC. LDA+U (U$_{eff}$=2.04 eV) points to a FM ground state and the importance of electron correlation effects. Using the Hubbard-corrected Generalized Gradient Approximation (GGA+U), it has been shown that Cr$_{2}$GeC is a weak AFM material for different exchange-correlation functionals \cite{Maurizio}. Cr$_{2}$GeC has similar degenerated magnetic states as Cr$_{2}$AlC that has been predicted to have FM ordering \cite{Ramzan1,Ramzan2}. However, that is opposite to the AFM coupling, predicted by Zhou \textit{et al.} \cite{Zhou}. Experimentally, Cr$_{2}$AlC and Cr$_{2}$GeC have been found to be FM at very low temperature (2.2 K) and very high external magnetic fields (10 T) \cite{Jaouen}. 

In this paper, we investigate the magnetism in a high-quality single-crystal Cr$_{2}$GeC (0001) thin film sample, using element-specific, polarization-, and temperature-dependent magnetic circular dichroism (XMCD)\cite{Mn-pek}. By changing the circular polarization from right to left-handed, and the direction of the applied magnetic field, the magnetic Cr moments are probed. We provide direct evidence of predominant ferrimagnetic ordering in the electronic structure supported by \textit{ab initio} band structure calculations. The measurements and calculations indicate a competition between a FM ordering in the basal $ab$-plane and an AFM spin distribution along the $c$-axis in the hexagonal crystal lattice.
\section{Experimental and computational details}

\subsection{Cr$_{2}$GeC (0001) thin film synthesis}
Phase-pure Cr$_{2}$GeC (0001) thin films were deposited at 800 $^o$C by dc magnetron sputtering from elemental Cr (99.95\% purity), Ge (99.999\%), and graphite (99.999\%) targets in an argon discharge pressure of 4 mTorr on MgO (111) substrates. The substrates were ultrasonically degreased in acetone followed by ethanol for five minutes and then annealed in vacuum in the deposition chamber for 1 h at 800 $^o$C prior to deposition. The Ar sputtering gas had a pressure of $\sim$0.3 Pa, and the base pressure was below 5 $\times$ 10$^{-6}$ Pa. The targets (75 mm diameter) were arranged on a confocal magnetron cluster and located at a distance of 18 cm from the substrate, which was mounted on a rotating sample holder. The Cr and C targets were run in DC power-control mode, while the Ge target was RF sputtered. The films were deposited to a thickness of $\sim$ 190 nm, corresponding to a deposition time of 30 min. Details on the synthesis process are given in Refs. \cite{Eklund2,Magnuson0}. 

\subsection{X-ray magnetic circular dichroism measurements}
The XAS and XMCD spectra were measured in total electron yield (TEY) mode with 0.15 eV energy resolution at 45$^{o}$ incidence angle on beamline I1011 at the MAX IV Laboratory \cite{Kovalik}. Left and right handed circularly polarized X-rays were provided in the 3rd harmonic with an elliptically polarizing undulator (EPU) and a collimated plane grating monochromator (cPGM) that enabled element specific characterization of magnetic properties of both ferromagnetic and anti-ferromagnetic materials. Constant magnetic fields of 0.2 and 0.4 T were applied both in-plane and out-of-plane on the sample with an octupole magnet both at room temperature and at 44 K.

\subsection{Computational details}

\subsubsection{Electronic ground state and magnetic spin coupling}
The electronic structure and magnetic ordering in Cr$_{2}$GeC was studied with the {\sc wien2k} code \cite{wien2k} employing the density-functional \cite{dft1,dft2} augmented plane wave plus local orbital (APW+lo) computational scheme. The Kohn-Sham equations were solved by means of the Wu-Cohen GGA ($GGA-WC$) \cite{GGA_WC1,GGA_WC2} for the exchange-correlation ($xc$) potential. 
The well-known shortcoming of the DFT description of the electron-correlation effects was explicitly treated within a phenomenological many-body Hamiltonian, the Hubbard model \cite{Hubbard}, where the effective on-site Coulomb interaction (U$_{eff}$) has been calculated
\cite{Maurizio} for the Cr atom in the hexagonal Cr$_{2}$GeC structure by using the constrained DFT formalism method \cite{Anisimov1991}.
A plane-wave expansion with R$_{MT}$$\cdot$K$_{max}$=10 was used in the interstitial region, while the potential and the charge density were Fourier expanded up to G$_{max}$=12. The modified tetrahedron method \cite{Blochl} was applied to integrate inside the Brillouin zone,
and a {\bf k}-point sampling with a 35$\times$35$\times$7 Monkhorst-Pack \cite{Monkhorst} mesh in the full BZ (corresponding to 786 irreducible {\bf k}-points) was considered satisfactory for the hexagonal Cr$_{2}$GeC system. Relativistic corrections ($e.g.$, spin-orbit coupling) in the electronic 
structure calculation were included in a second-variational procedure using scalar relativistic wave functions \cite{Singh1994}.
The magnetic ground-state properties were studied after having achieved the relaxed unit cell parameters ($a$=2.981 \AA{} and $b$=12.044  \AA{}). 

The computed $GGA-WC+U$ magnetic spin ordering in Cr$_{2}$GeC is generally AFM along the out-of-plane $c$-axis with residuals of FM within the $ab$ basal plane. 
The details of this magnetic spin coupling were presented in an earlier work \cite{Maurizio}, where a Ge-mediated super-exchange magnetic
 structure interact between the piled CrC layers (Fig. 1, bottom panel).

To test the robustness of this magnetic ground-state scenario, we further applied different theoretical schemes. This included a pseudopotential calculational method ($q$-ESPRESSO \cite{QE}) based on a hybrid exchange-correlation functional carried out on a 2$\times$2$\times$1 supercell (B3LYP\cite{B3LYP}), coupled with a non spin collinear treatment of magnetism. The results showed that there is still a general AFM spin distribution, although a non-vanishing non collinear magnetic component is evidenced both along the $a$- (0.02 $\mu_{B}$/cell) and $c$-axis (0.05 $\mu_{B}$/cell). 

\subsubsection{Theoretical XMCD spectra}
The absorption cross-section ($\mu$) for incident X-rays was theoretically determined according to Fermi's golden rule by using the one-particle framework and the dipole approximation by the probability of an electron to be excited from a core state to a final valence state. Our dichroic and total absorption spectra were computed within the {\sc wien2k} code \cite{wien2k} through the linear combinations of the absorption cross-sections ($\mu^{+}$-$\mu^{-}$ and $\mu^{+}$+$\mu^{-}$), by using the LAPW basis set formalism of Pardini {\it et al.} \cite{Pardini}. Convergent theoretical XMCD spectra were achieved within the same $GGA-WC+U$ methodology presented earlier in ref. \cite{Maurizio}. Reference XMCD spectra were also computed within the Green formalism ($i.e.$, multiple scattering) on a muffin-tin potential \cite{MT} implemented in the FDMNES package \cite{FDMNES}. Well-converged XMCD spectra were obtained by using a cluster's radius of 8 \AA. Such a less precise but computationally faster methodology was systematically applied on a variety of AFM and FM spin configurations in order to get a deeper insight on the type of magnetic coupling in Cr$_{2}$GeC.

\section{Results}

\begin{figure}
\includegraphics[width=80mm]{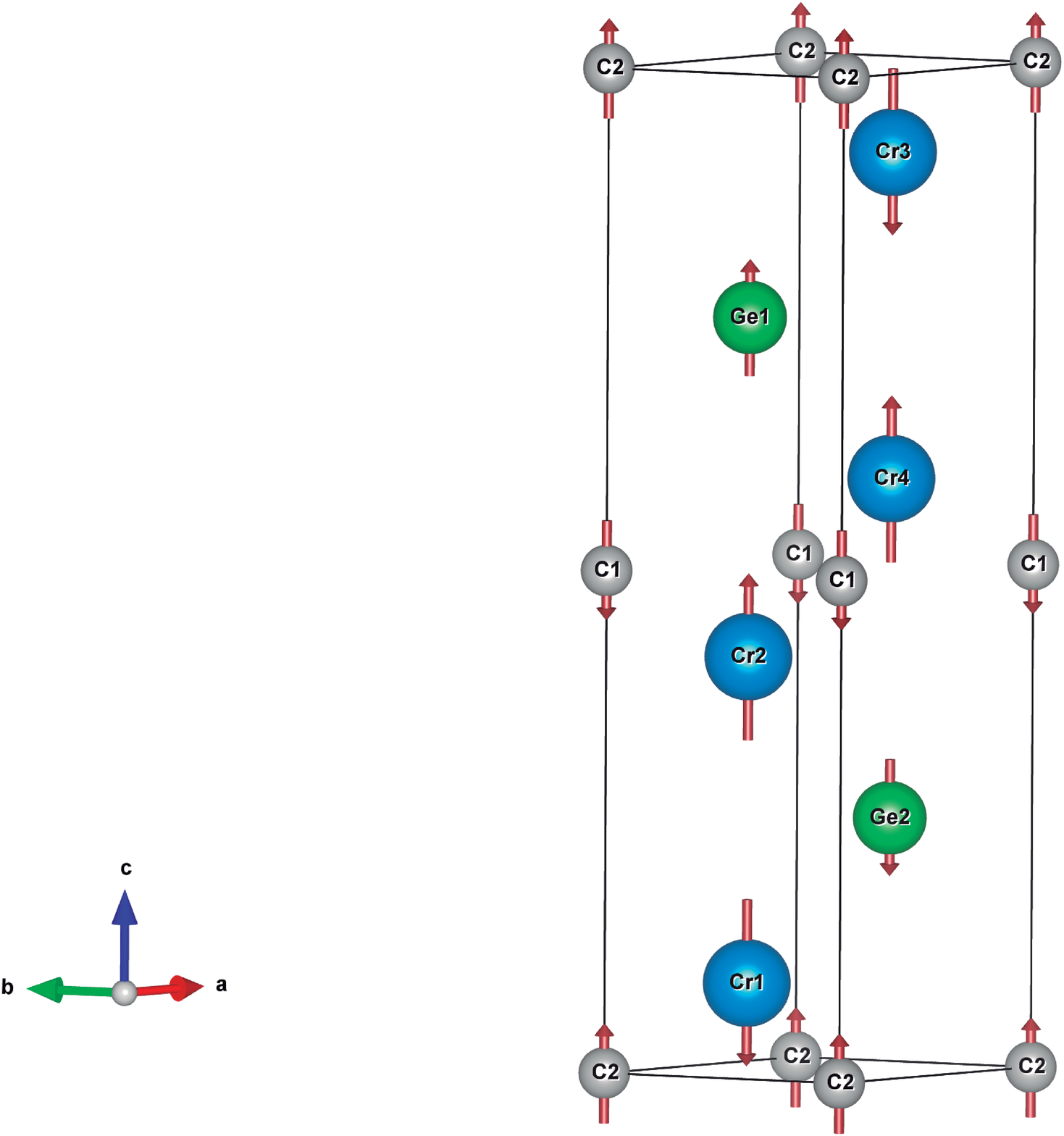} 
\includegraphics[width=80mm]{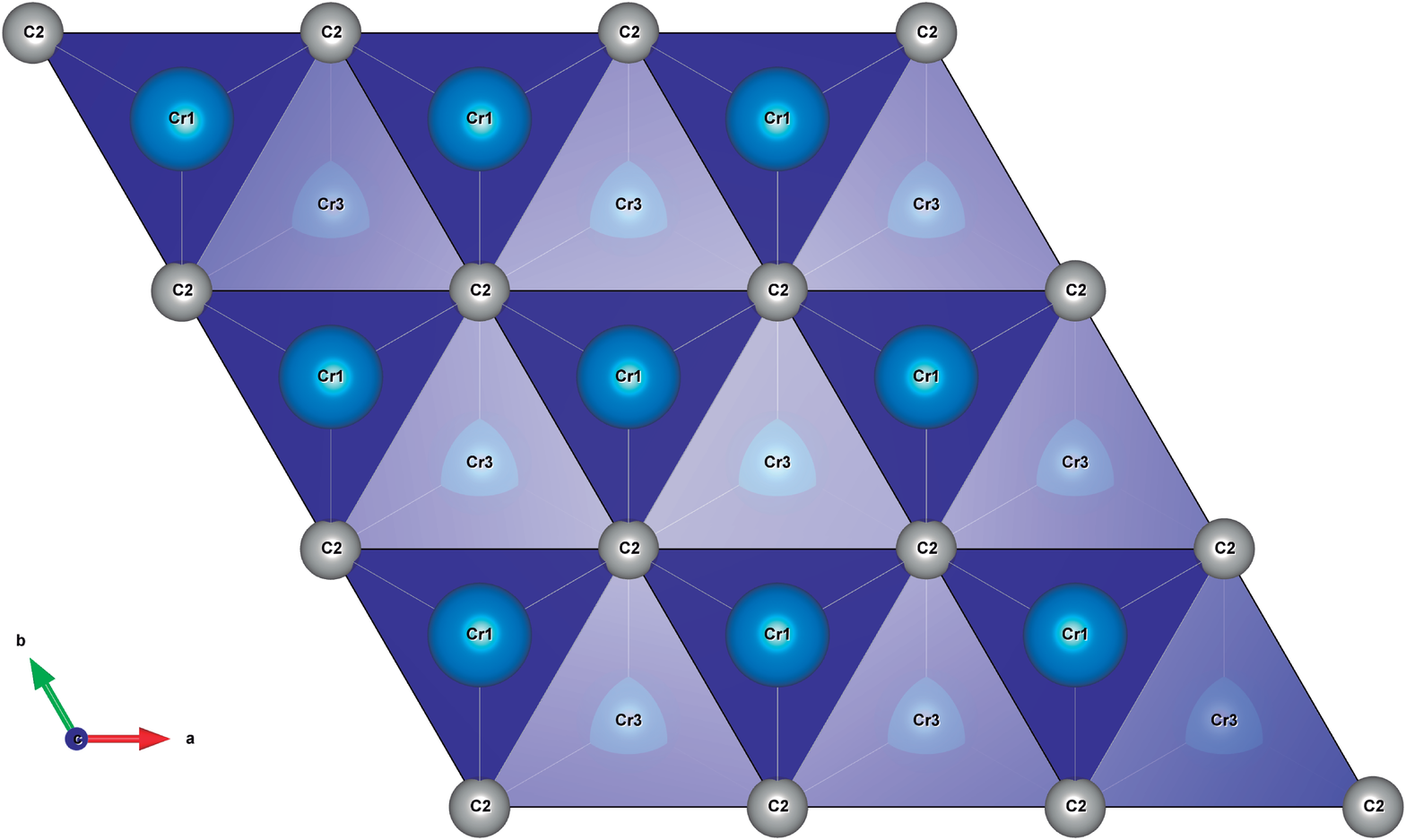} 
\vspace{0.2cm} 
\caption[] {(Color online) Top panel: Illustration of the hexagonal crystal structure of Cr$_2$GeC with atoms of different spins.
The CrC slabs are interleaved by pure layers of Ge. Bottom panel: Schematic representation of the CrC polyhedra of the laminate plane.}
\label{structure}
\end{figure}

Figure 1 shows the unit cell of Cr$_{2}$GeC with a hexagonal crystal structure containing CrC slabs interleaved by atomic layers of Ge atoms. 
Calculated magnetic moments are indicated by the arrows where the lengths correspond to the relative size of the moments and the direction correspond to the spin orientation obtained from the results of $GGA-WC+U$ calculations \cite{Maurizio}.
The unit cell shows alternating Cr-C slabs with an in-plane FM Cr-Cr coupling where the Cr-containing basal $ab$-planes are piled up into a nanolaminate along the $c$-axis with an AFM spin ordering.

\begin{figure}
\includegraphics[width=75mm]{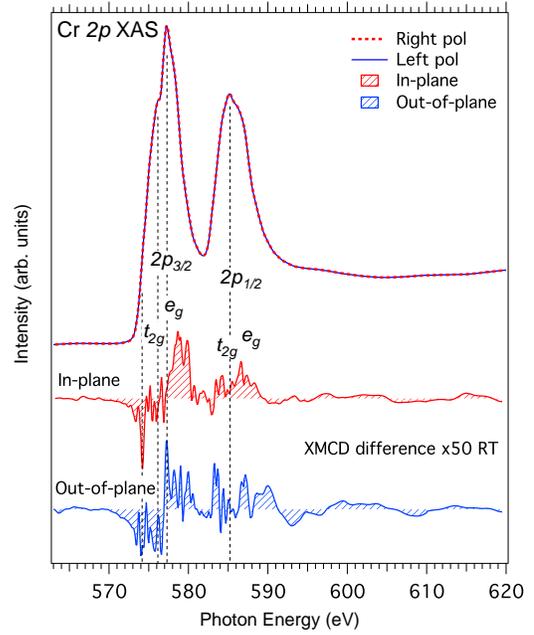} 
\vspace{0.2cm} 
\caption[] {(Color online) Experimental Cr $2p$ X-ray absorption spectra with left and right hand polarization in a magnetic field of 0.4 T at room temperature. Bottom: XMCD difference spectra measured with the magnetic field in the basal $ab$-plane and out-of-plane along the $c$-axis. }
\label{Cr 2p}
\end{figure}

Figure 2 (top) shows Cr $2p_{3/2,1/2}$ X-ray absorption spectra (XAS) measured at room temperature with left- and right-hand polarization in a constant magnetic field of 0.4 T. XMCD difference spectra are shown below for in-plane and out-of-plane alignment of the magnetic field. Although the general shapes of the XAS spectra appear similar for the two orientations of the magnetic field, the XMCD spectra exhibit significant differences. 
The observed sub-peak splitting of the $2p_{3/2}$ and $2p_{1/2}$ peaks in XAS is due to the $t_{2g}$ and $e_{g}$ splitting that are salient features for the metallic and bonding orbitals, respectively. The $t_{2g}$($d_{xy}$, $d_{xz}$, $d_{yz}$) orbitals have the lowest energy while the covalent $e_{g}$($d_{3z^2-r^2}$, $d_{x^2-y^2}$) orbitals are located at 1.2 eV higher energy in the hexagonal symmetry surrounding the Cr sites. 
The $t_{2g}$-$e_{g}$ splitting at the $2p_{3/2}$ and $2p_{1/2}$ peaks mainly affects the shape of the XAS but not necessary the shape of the XMCD signal. The latter depends primary on the difference between left- and right-hand polarized edges.
The $2p_{3/2}$/$2p_{1/2}$ branching ratio of 1.27 is smaller than the statistical ratio of 2:1 and reflects the amount of conductivity, the exchange and mixed terms between the core-states \cite{Laskowski}. 

The observed intensity oscillations of the XMCD signal at the $2p_{3/2}$-edge of Cr$_{2}$GeC with negative intensity at $\sim$574 eV followed by positive intensity at $\sim$579 eV indicate that the magnetism in Cr$_{2}$GeC consists of a mixture of FM and AFM coupling of the spins of the Cr $3d$ electrons. 
In the basal $ab$-plane, the XMCD signal of Cr show similarities to induced magnetism of Cr $3d$ states in Fe/Cr multilayers \cite{Tomaz,Alayo}. However, along the $c$-axis, the spectral shape exhibits a more randomly ordered distribution of negative and positive intensity that is a sign of a predominant AFM ordering. 

\begin{figure}
\includegraphics[width=75mm]{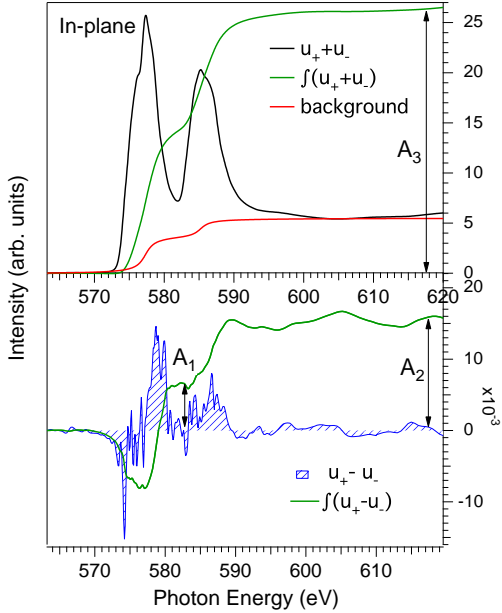} 
\vspace{0.2cm} 
\caption[] {(Color online) In-plane analysis of experimental Cr $2p$ x-ray absorption spectra with left and right hand polarization in a magnetic field of 0.2 T at 44 K. Bottom: XMCD difference spectra measured with the magnetic field in the basal $ab$-plane. The A$_{1}$, A$_{2}$ and the A$_{3}$ represent the three integrated areas needed in the sum-rule analysis.}
\label{XMCD1}
\end{figure}

Figure 3 shows XAS and XMCD data measured at 44 K with an external field of 0.2 T in the basal $ab$-plane. By applying LHe temperatures, the spectral features become slightly sharper than in the room temperature measurements in Fig. 2. In order to quantitatively estimate the magnetic moments in the basal $ab$-plane, the spectra have been analyzed using the \textit{sum rules} that relate the integrated XMCD intensities to the element-specific properties of the $3d$ orbital angular momentum $m_{orb}$=-4$\times$$n_{h}$$\times$A$_{2}$/3A$_{3}$ and the $3d$ spin angular momentum $m_{spin}$=-$n_{h}$$\times$(6A$_{1}$-4A$_{2}$)/A$_{3}$ as originally suggested by Thole \textit{et al.} \cite{Thole}. In this way, the spin and orbital magnetic moments were determined from the integrated areas (denoted A$_{1}$, A$_{2}$ and the A$_{3}$ in Figs. 3 and 4) of the XMCD signal as experimentally confirmed for Fe and Co by Chen \textit{et al.} \cite{Chen}. As observed in the bottom panel of Fig. 3, the integrated XMCD curve crosses the zero line and a possible mixture of FM/AFM is indicated by the integrated XMCD signal ($u_{+}$ + $u_{-}$) that is first negative, crosses zero and then becomes positive. 

From the $GGA-WC+U$ calculations, we obtained a total valence charge of 5.55 electrons within the Cr atomic spheres ($n_{3d}$=1.60, $n_{3p}$=2.93 and $n_{spin}$=1.03, $n_{h}$=4.45). The A$_{1}$, A$_{2}$ and A$_{3}$ notations in Fig. 3 represent the three integrated areas needed in the sum-rule analysis, where A$_{3}$ is the integral for the whole $2p_{3/2,1/2}$ range that can be precisely determined. 
The separate A$_{1}$ and A$_{2}$  integrals of the $2p_{3/2}$ and $2p_{1/2}$ ranges depend on the cutoff energy and the potential orbital overlap between the edges. The orbital and spin moments were then determined by applying $n_{h}$=4.45 and the A$_{1}$, A$_{2}$ and the A$_{3}$ values indicated in Fig. 3 and, after correcting for the geometry of the beam and the polarization rate by 1/cos(45$^{o}$)/0.95 as, $m_{l}$=(5.3$\pm$0.5)$\cdot$10$^{-3}$$\mu_{B}$/atom and $m_{s}$=(6.4$\pm$1.0)$\cdot$10$^{-3}$$\mu_{B}$/atom.

\begin{figure}
\includegraphics[width=75mm]{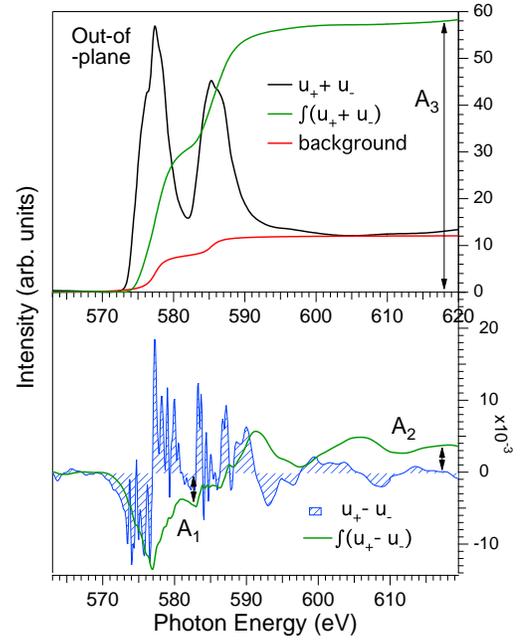} 
\vspace{0.2cm} 
\caption[] {(Color online) Out-of-plane analysis of experimental Cr $2p$ x-ray absorption spectra with left and right hand polarization in a magnetic field of 0.2 T at 44 K. Bottom: XMCD difference spectra measured with the magnetic field out-of-plane along the $c$-axis. The A$_{1}$, A$_{2}$ and the A$_{3}$ represent the three integrated areas needed in the sum-rule analysis.}
\label{XMCD2}
\end{figure}

Figure 4 shows XAS and XMCD data measured at 44 K with an external field of 0.2 T applied along the $c$-axis (out-of-plane). By applying the sum rules in the same way as in Fig. 3, we find $m_{l}$=(5.4$\pm$0.5)$\cdot$10$^{-4}$$\mu_{B}$/atom and $m_{s}$=(3.6$\pm$0.7)$\cdot$10$^{-4}$$\mu_{B}$/atom. 
Comparing the XMCD results in Figs. 3 and 4, the orbital and spin moments are an order of magnitude smaller out-of-plane than in-plane. The $m_{l}$ and $m_{s}$ values are similar in-plane but for out-of-plane, the spin moment $m_{s}$ is smaller than the orbital moment $m_{l}$. 

\begin{figure}
\includegraphics[width=75mm]{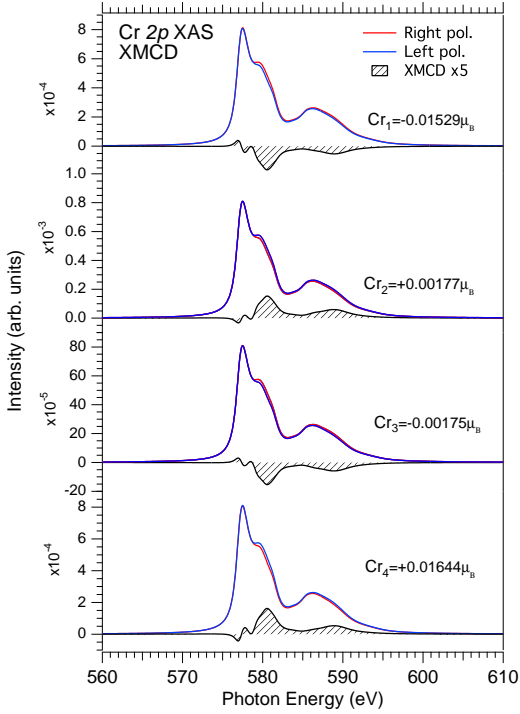} 
\vspace{0.2cm} 
\caption[] {(Color online) Calculated Cr $2p$ XAS and XMCD spectra for the four Cr atoms in Cr$_{2}$GeC with a spin-collinear treatment.
The shape and amplitude of the curves depends on the sign and strengths of the Cr moments.}
\label{XMCD3}
\end{figure}

Figure 5 shows the calculated atom-specific XAS and XMCD spectra using $GGA-WC+U$ with relativistic corrections for the four different Cr atoms in the Cr$_2$GeC unit cell. Note that in this type of calculations, even though intensities and spectral shapes of the XMCD spectra are rather similar, the signs of the magnetic moments are different. The largest moments were found for Cr$_{1}$ and Cr$_{4}$ but with opposite signs. The moments of Cr$_{2}$ and Cr$_{3}$ are $\sim$ 10 times smaller than those of Cr$_{1}$ and Cr$_{4}$ also with different signs. Such a magnetic coupling scenario, where small spin moment magnitudes are involved in a sort of entangled competition between FM and AFM coupling complicates the physical interpretation of the measured spectra. However, by looking at the sum of the XMCD spectra of Cr$_{2}$ and Cr$_{4}$ (\textit{i.e.}, the in-plane atoms coupled FM) we find that the in-plane XMCD is a non-zero signal, while the sum of Cr$_{1}$ + Cr$_{2}$ and Cr$_{3}$ + Cr$_{4}$ give almost a null XMCD (\textit{i.e.}, an AFM coupling along the $c$-axis). Thus, the signal originating from the sum of Cr$_{2}$ and Cr$_{4}$ will be used (Fig. 6, orange line) to compare with the experimental XMCD spectrum.
In order to further understand the nature of magnetism in Cr$_{2}$GeC, reference XMCD model spectra for frozen FM and AFM configurations were calculated within the faster Green formalism. 

\begin{figure}
\includegraphics[width=75mm]{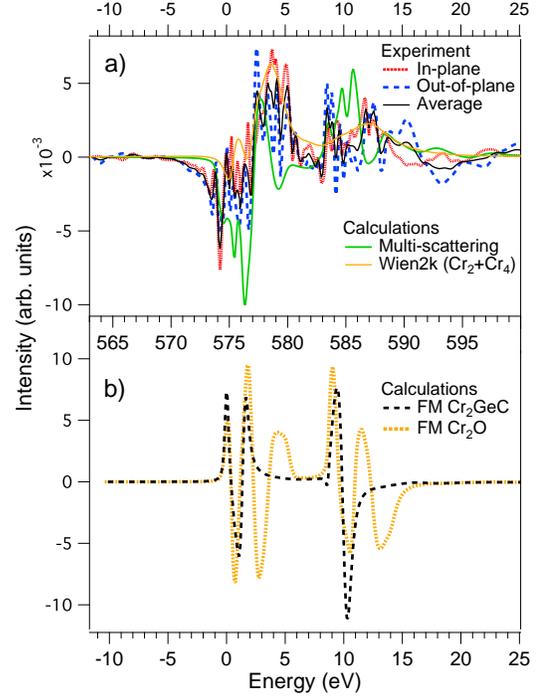} 
\vspace{0.2cm} 
\caption[] {(Color online) The upper panel (a) shows the experimental Cr $2p$ XMCD spectra of Figs. 3 and 4 in comparison to model spectra the with same FM/AFM configuration as shown in Fig. 1. The green curve use multi-scattering formalism, while the orange curve refers to the Cr$_{2}$+Cr$_{4}$ spectral shape using the $GGA-WC+U$ method. Bottom panel (b) shows calculated reference XMCD spectra for a pure FM spin coupling in both Cr$_{2}$GeC and CrO$_{2}$. }
\label{XMCD3}
\end{figure}

Figure 6 (top panel) shows the experimental XMCD spectra of Figs. 3 and 4 in comparison to model spectra with the same spin coupling distribution in the Cr$_{2}$GeC as shown in Fig. 1 using both the $GGA-WC+U$ and the full-multi-scattering methods. Although the Cr$_{2}$+Cr$_{4}$ spectral shape is not in perfect quantitative agreement with the measured spectrum, we note that qualitatively, this calculation catch up the positive part of the high energy XMCD signals.
Fig. 6 (bottom panel), shows the CrO$_{2}$ reference spectrum in comparison to that of a pure FM scenario in Cr$_{2}$GeC.
Although the fitting is not perfect, the multi-scattering method (green line in Fig. 6) shows that a scenario with FM coupling in the basal $ab$-plane and AFM along the $c$-axis of Cr$_{2}$GeC (\textit{i.e.}, that of Fig. 1), is clearly in much better agreement than a pure FM coupling in both directions. In particular, the first and second peaks around 573.2 eV and $\sim$ 574 eV in pure FM coupling have opposite signs in comparison to the experimental results as well as for the calculated orthogonal FM/AFM coupling in Cr$_{2}$GeC. This is also the case for a possible CrO$_{2}$ contribution on the surface, as the shape of the XMCD spectrum is significantly different from the experimental results and thus unlikely (bottom panel of Fig. 6). To further test this scenario, several other 2$\times$2$\times$2 supercells were used to test different AFM/FM spin couplings within the same Green formalism and found that only the spin ordering shown in Fig. 1 provides sufficient agreement with the experimental data.
Furthermore, we note that the full-multi scattering method is generally in better agreement with measurements at the low-energy part of the spectrum, while at higher energies, the $GGA-WC+U$ method performs better. This behavior is in line with the fact that the more computational demanding $GGA-WC+U$ calculations were performed without considering the core-hole effect, while in the multi-scattering scheme, the $\it{Z+1}$ approximation was used to take this interaction into account. On the other hand, the Green formalism shows a certain shortcoming description for the energy position of positive peaks located at higher energies.  This behavior can be addressed to the well-known scissor operator effect \cite{Sole} and to the calculated $first-principles$ values of the spin-orbit splitting that are slightly underestimated for the early transition metals \cite{Review}.

\begin{table*}[tp]
\caption[tabbetas]{\label{tabII}\sf Calculated ($GGA-WC+U$ with relativistic corrections) magnetic moments (in $\mu_{B}$/cell), and their orbital ($m_{l}$) and spin ($m_{s}$) contributions.}
\[\begin{array} {l c c c}
\hline
\multicolumn{1}{c}{\mbox{Atom}}&\multicolumn{1}{c}{\mbox{spin}} &\multicolumn{1}{c}{\mbox{$m_{l}$}}&\multicolumn{1}{c}{\mbox{$m_{s}$}}\\
\hline
{\mbox{Cr$_{1}$}} &-0.01529  &-0.47522  &-0.01404\\
{\mbox{Cr$_{2}$}} &+0.00177  &-0.47562  &+0.00181\\
{\mbox{Cr$_{3}$}} &-0.00175  &+0.47580  &-0.00238\\
{\mbox{Cr$_{4}$}} &+0.01644  &+0.47535  &+0.01510\\
\hline
\end{array}\]
\end{table*}

\begin{table*}[tp]
\caption[tabbetas]{\label{tab:sml.1}\sf Calculated (B3LYP+non-spin collinear) in-plane vector components ($m_{x}$, $m_{y}$) and out-of-plane ($m_{z}$) magnetic moments (in $\mu_{B}$/cell) of the Cr, Ge and C atoms of Cr$_{2}$GeC.}
\[\begin{array} {l c c c }
\hline
\multicolumn{1}{c}{\mbox{Atom}}&\multicolumn{1}{c}{\mbox{$m_{x}$}}&\multicolumn{1}{c}{\mbox{$m_{y}$}}&\multicolumn{1}{c}{\mbox{$m_{z}$}}\\
\hline
{\mbox{Cr$_{1}$}} &{\mbox{+4.58$\cdot$10$^{-3}$}} &{\mbox{+3.80$\cdot$10$^{-5}$}} &-2.81  \\
{\mbox{Cr$_{2}$}} &{\mbox{+4.56$\cdot$10$^{-3}$}} &-{\mbox{4.50$\cdot$10$^{-5}$}} &+2.79 \\
{\mbox{Cr$_{3}$}} &{\mbox{+4.86$\cdot$10$^{-3}$}} &{\mbox{+3.20$\cdot$10$^{-5}$}} &-2.81 \\
{\mbox{Cr$_{4}$}} &{\mbox{+4.98$\cdot$10$^{-3}$}} &-{\mbox{3.10$\cdot$10$^{-5}$}} &+2.79 \\
{\mbox{Ge$_{1}$}} &-{\mbox{1.21$\cdot$10$^{-4}$}} &0.0 &-{\mbox{4.14$\cdot$10$^{-3}$}}\\
{\mbox{Ge$_{2}$}} &-{\mbox{1.25$\cdot$10$^{-4}$}} &0.0 &-{\mbox{2.37$\cdot$10$^{-3}$}}\\
{\mbox{C$_{1}$}} &-{\mbox{1.04$\cdot$10$^{-3}$}} &0.0 &+0.61\\
{\mbox{C$_{2}$}} &-{\mbox{1.06$\cdot$10$^{-3}$}} &0.0 &-0.61\\
\hline
{\mbox{$m_{tot}$}}&{\mbox{+1.66$\cdot$10$^{-2}$}} &{\mbox{$\sim$0}} &{\mbox{-6.50$\cdot$10$^{-3}$}}\\
\hline
\end{array}\]
\end{table*}

Table I shows the calculated self-consistent magnetic moments for Cr$_{2}$GeC (in $\mu_{B}$/cell), and their orbital ($m_{l}$) and spin ($m_{s}$) contributions using $GGA-WC+U$ with relativistic corrections. The average spin moments are $<$m$_{l}$$>$=7.75$\cdot$10$^{-5}$$\mu_{B}$ and $<$m$_{s}$$>$=1.23$\cdot$10$^{-4}$$\mu_{B}$. In general, the computed results with small magnetic moments are in agreement with the experimental AFM coupling although the calculated $m_{s}$ value is somewhat smaller than the experimental values obtained from the spin sum rule. Note that for weak AFM materials such as Cr$_{2}$GeC, the orbital moments $m_{l}$ are often more important than the spin moments $m_{s}$ in XMCD. The values of the pairs of orbital moments basically have the same magnitudes, and only a change in sign is observed for Cr$_{1}$ and Cr$_{4}$ as well as for Cr$_{2}$ and Cr$_{3}$ (Table I). Experimentally, we further observe that the average spin moments are an order of magnitude larger in-plane compared to out-of-plane.

In order to shed more light onto the difference between in-plane vs. out-of-plane spin distributions, we also carried out hybrid-functional calculations using a non-collinear spin-treatment. Table II is the result of a different calculation scheme based on B3LYP+non-collinear treatment that yield different results in terms of total magnetic moments. The results obtained by using the hybrid-functional methodology shows that each magnetic moment of Cr, Ge and C atoms in Cr$_{2}$GeC can be decomposed along the laminated $ab$-plane ($m_x$, $m_y$) and along the $c$-axis ($m_z$). The major contributions to the spin appear along the $c$-axis ($m_z$) of Cr while the small magnetic moments of C and Ge elements are due to induced polarization effects. The experimental observations are in good agreement with the results achieved within the non-collinear spin treatment of the Cr$_2$GeC system (Table II), where the obtained in-plane net magnetic moment (1.66$\cdot$10$^{-2}$$\mu_{B}$) is 2.6 times larger than the out-of-plane (6.50$\cdot$10$^{-3}$$\mu_{B}$). Although a common AFM ordering is predicted along the $c$-axis in both cases, the calculations indicate that electronic correlation effects are critical in describing the correct magnetic spin ordering in Cr$_{2}$GeC that will be further discussed below.

\section{Discussion}
In the XMCD experiments shown in Figs. 3 and 4, the net AFM coupling observed by the average Cr atoms in- and out-of-plane is a result of the weighted coupling from the Cr-Cr, Cr-C and Cr-Ge interactions. Thus, the net magnetic Cr coupling not only depends on the magnitudes and distances between the FM coupled Cr atoms, but also, on the AFM coupled C atoms that are closer to the Cr as shown in Fig. 1.  

For the quantitative determinations of the average magnetic moments of Cr, the limitations of the integral sum rules should be considered \cite{Scherz}.  In particular, the spin moment is sensitive to the number of $3d$ holes and the Cr $L_{3}$/$L_{2}$ branching ratio.
This depends on the Cr $3d$-band filling, the amount of overlap of the Cr $2p_{3/2,1/2}$ edges and the Cr $2p_{3/2}$/$2p_{1/2}$ branching ratio.
This is more severe for the earliest transition metals Ti and V where the $2p$ spin-orbit splitting is smaller and imply more mixing of the $2p_{3/2}$ and $2p_{1/2}$ core states \cite{Stohr1}. 
Moreover, the spin sum rule $<$m$_{s}$$>$ that involves the difference between the integrated areas at the $2p_{3/2}$ and $2p_{1/2}$ edges is most sensitive to a change of the $2p_{3/2}$/$2p_{1/2}$ branching ratio.
Therefore, the $<$m$_{s}$$>$ spin moments in the sum rules are often underestimated while the orbital moments $<$m$_{l}$$>$ usually yield reasonable values.

XMCD measurements on MAX-phases with hexagonal symmetry are thus complex, not only requiring sample phase purity but also avoiding surface oxides that quickly form on fresh surfaces in air. 
Although Cr$_{2}$GeC is oxidation resistant and the sample was properly cleaned and ultrasonically degreased in acetone followed by ethanol before being measured in ultrahigh vacuum, the formation of a thin natural nonmagnetic Cr$_{2}$O$_{3}$ or ferromagnetic CrO$_{2}$ oxides on the surface cannot be completely excluded.
In addition, the very weak magnetic signals require extensive measurement times in stable X-ray beams and magnetic fields. Furthermore, the nature of the correlation effects of localized Cr $3d$ states and the competing balance between FM and AFM ordering with a very small energy difference, that makes modeling of the magnetic coupling in Cr$_{2}$GeC rather complicated. 
If the non-collinearity of the spins is fully taken into account in Cr$_{2}$GeC, the presence of two competing Cr-Cr magnetic mechanisms (FM in plane and AFM along the $c$-axis), leads to stabilization of a non-perfectly compensated AFM material. Indeed, by employing DFT+U full-potential calculations with different exchange-correlation functionals, Mattesini \textit{et al.} \cite{Maurizio} already showed that Cr$_{2}$GeC is a weak AFM material. For this purpose, the Cr$_{2}$GeC phase can be used as a model test case for other similar magnetic MAX phases such as, for example, Cr$_{2}$AlC \cite{Schneider}, and V$_{2}$GeC \cite{Luo}.

Although the Cr $3d$ states theoretically exhibit a topologically FM coupling within the $ab$-planes, the net magnetic moment of Cr$_{2}$GeC in a unit cell is close to zero when these planes piles-up anti-ferro-magnetically along the $c$-axis. Comparing the results shown in Tables I and II, they are both in general agreement with an overall AFM ordering and provide the same magnetic scenario found experimentally through the XMCD spectra. The difference is merely on the magnitude of the localized magnetic moments of Cr. Although the hybrid B3LYP functional coupled to a non-spin collinear treatment provides Cr magnetic moments that are larger than what is found within the $GGA-WC+U$ \cite{Maurizio}, both methods point to the same vertical AFM spin ordering for the Cr$_{2}$GeC unit cell. 

In addition to the AFM coupling between Cr atoms of different $ab$-basal planes, as discussed above, there is, a small but non-negligible spin-collinear component affects the magnetic properties. The non-spin collinear component influences the stacking of ferromagnetic layers with spins inside the layers that are not perfectly vertically aligned, resulting in a non-compensated antiferromagnetic coupling along the $c$-axis. Thus, it is the presence of a weak spin-wave effect that makes the entire system a magnetically strained AFM material, where the final macroscopic magnetic properties depend on small changes in the synthesis conditions $i.e.$, temperature and sputtering flux. We suspect that this is one of the reasons why the experimental characterization of the magnetic properties of Cr$_{2}$GeC has been so far rather controversial \cite{Maurizio,Ramzan1,Ramzan2,Jaouen,Zhou}. 

The competition between FM and AFM coupling is referred to as antiferromagnetic spin frustration \cite{Grossi}. We anticipate that topological constraints from the hexagonal crystal lattice should be avoided in order to stabilize a FM coupled pure MAX-phase. 
One solution to overcome this is to break the ordering symmetry and achieve configuration disorder by alloying and replacing part of the metal atoms with other magnetic elements such as Fe and Mn \cite{Grossi}.
The observed low Curie temperature for both Cr$_{2}$GeC (0 K) \cite{Liu} and Cr$_{2}$AlC (73 K) \cite{Jaouen}, suggest that magnetic ordering should only be significant below T$_{c}$. However, as shown for (Cr$_{x}$,V$_{1-x}$)$_{2}$AlC phases in ref. \cite{Grossi}, this is not the case as NM calculations yield poor results in comparison to experiments. Nonetheless, it is reasonable to expect that for Cr$_{2}$GeC the observed net Cr FM moments are very small at room temperature \cite{Liu}. 

We further believe that the relatively small spin magnitudes of the C atoms (C$_{1}$ and C$_{2}$) shown in Table II stabilize the ferromagnetic coupling of the Cr atoms within each layer, while the individual spins of the Ge atoms (Ge$_{1}$ and Ge$_{2}$) are essential for establishing a Ge-mediated super-exchange coupling between the vertically piled layers. This interpretation is consistent with findings in other two-dimensional materials $e.g.$, metal-coordinated networks and ferrites \cite{Mn-pek,NA}. Considering all possibilities of spin distortions modulated by a 3D spin-wave distribution, the unit cell is thus not a perfect anti-ferromagnet with a null total spin but, instead exhibits residual magnetic components both along the $ab$-basal plane and along the $c$-axis with a total magnetization of 0.02 and 0.05 $\mu_{B}$/cell, respectively. Indeed, in recent $GGA-WC+U$ calculations of Cr$_{2}$GeC, Mattesini \textit{et al.} \cite{Maurizio} and Magnuson \textit{et al.} \cite{Magnuson0} pointed out that the magnetic moments
on the Cr atoms, although in-plane FM coupled, are small and largely cancel each other along the $c$-axis. The latter vertical coupling could lead to a perfect AFM ordering or to a ferrimagnetic ground-state if a non-spin collinear effect is taken into account. The Heyd-Scuseria-Ernzerh (HSE06) hybrid functional formalism \cite{Heyd} has been employed for a variety of different spin configurations and also provides an AFM ground state ordering \cite{Ramzan2,Sun}. Thus, in addition to the existing controversial theoretical interpretations about what type of magnetic coupling dominates ($i.e.$, in-plane or out-of-plane), the evident disagreement between theory (AFM) and experiment (FM) in determining the lower energy magnetic ordering in Cr$_{2}$GeC calls for the need of further experimental and theoretical efforts  \cite{Ramzan2,Maurizio} to disentangle the temperature dependence and the effect of phonons.

For a correct description of magnetic states and correlation effects in MAX-phases, calculations needs to go beyond the rigid band model.
In further studies of potential magnetic properties in MAX-phases, both different competing magnetic interactions and temperature effects should be included and carefully analyzed. 
The $t_{2g}$-$e_{g}$ branching ratio can also be improved by many-body perturbation theory by solving the Bethe-Salpether equation (BSE) implemented in WIEN2kbse \cite{Laskowski}. 
Phonons are also known to play a role in many of the physical properties of condensed matter, such as thermal and electrical conductivity and magnetism \cite{Kim}. Besides the phonon-induced electro-structural effects featured earlier in Cr$_{2}$GeC \cite{Magnuson0}, electron-phonon interactions can further influence the competition between AFM and FM coupling in Cr-based MAX-phases. In this sense, the Ge atoms in Cr$_{2}$GeC move preferentially within the $ab$-basal plane, while Cr and C have preference for the $c$-axis direction. As shown in previous phonon calculations \cite{Magnuson0}, a $x$-, $y$-displacement of 0.097 \AA{} and a larger $z$-displacement of 0.114 \AA{} was found for the Cr atoms. Rapidly moving atoms is known to change the electronic DOS at E$_{F}$ \cite{Magnuson0} and should also affect the magnetic ordering \cite{Kim}. This implies that phonons play an important role especially in strained magnetic systems. Specifically, the movement of Cr atoms can in different ways affect the spin ordering along the $c$-axis $vs$. the in-plane ordering. It can be expected that at room temperature experiments where the phonon amplitudes are larger than at 0 K, the detection of the in-plane FM coupling becomes more challenging. On the contrary, the AFM spin coupling along the $c$-axis should be less affected by thermally dependent atomic motions.

\section{Conclusions}

This work provides a step forward toward understanding the controversial magnetism in an important class of nanolaminated materials.
X-ray magnetic circular dichroism have been applied to investigate the complicated magnetic coupling mechanism in the Cr$_{2}$GeC MAX-phase. 
The measurements exhibit a predominantly antiferromagnetic coupling between the Cr layers along the $c$-axis that is influenced by a ferromagnetic contribution in the nanolaminated $ab$-basal planes. 
Experimentally, we also find that the net magnetic moments along the $ab$-basal plane is ten times larger than along the $c$-axis. 
We showed that this results in an overall residual spin moment that resembles that of a ferrimagnet. $Ab$ $initio$ calculations further confirm that Cr$_{2}$GeC is a magnetically strained system that exhibits small but different resultant magnetic components along the $ab$-basal plane and along the $c$-axis.

\section{Acknowledgements}

We thank the staff at MAX IV Laboratory for experimental support. 
We thank P. Eklund, M. Bugnet and V. Mauchamp for providing the samples and M. Jaouen for discussions. 
This work was financially supported by the Swedish Research Council, Linnaeus Grant LiLi-NFM and the 
the Swedish Foundation for Strategic Research (SSF). M. Magnuson acknowledges financial support from the Swedish Energy Research (no. 43606-1) and the Carl Tryggers Foundation (CTS16:303, CTS14:310). 
M. Mattesini acknowledges financial support by 
the Spanish Ministry of Economy and Competitiveness (CGL2013-41860-P and CGL2017-86070-R).


\begin{thebibliography}{100}

\bibitem{Barsoum1} M. W. Barsoum; The M$_{N+1}$AX$_{N}$ phases: A new class of solids: Thermodynamically stable nanolaminates,  \textit{Prog. Solid State Chem.} \textbf{2000} 28, 201-281.

\bibitem{Wang}  Wang, J.; Zhou, Y., Recent progress in theoretical prediction, preparation, and characterization of layered ternary transition-metal carbides,  \textit{Annu. Rev. Mater. Res.} \textbf{2009}, 39, 415-443.

\bibitem{Eklund1} Eklund, P.; Beckers, M.; Jansson, U.; H\"{o}berg, H.; Hultman, L.; The M$_{n+1}$AX$_{n}$ phases: Materials science and thin-film processing,  \textit{Thin Solid Films} \textbf{2010} 518, 1851-1878.

\bibitem{Review} Magnuson, M.; Mattesini, M.; Chemical bonding and electronic-structure in MAX phases as viewed by X-ray spectroscopy and density functional theory,  \textit{Thin Solid Films} \textbf{2017} 621, 108-130.

\bibitem{Barsoum_book} Barsoum, M. W., \textbf{2013} \textit{MAX Phases: Properties of Machinable Ternary Carbides and Nitrides}, Wiley.

\bibitem{Ahuja} Luo, W.; Ahuja, R., Magnetic Fe$_{n+1}$AC$_{n}$ (n = 1, 2, 3, and A = Al, Si, Ge) phases: from ab initio theory,  \textit{J. Phys. Cond. Mat.} \textbf{2008} 20, 064217.

\bibitem{Dahlqvist} Dahlqvist, M.; Alling, B.; Abrikosov, I. A.; Rosen, J., Effect of magnetic disorder and strong electron correlations on the thermodynamics of CrN,  \textit{Phys. Rev. B} \textbf{2010} 82, 184430.

\bibitem{Lin} S. Lin, P. Tong, B. S. Wang, Y.N. Huang \textit{et al.}; Magnetic and electrical/thermal transport properties of Mn-doped M$_{n+1}$AX$_{n}$ phase compounds Cr$_{2-x}$Mn$_{x}$GaC (0$\leq$x$\leq$1),  \textit{J. Appl. Phys.} \textbf{2013} 113, 053502.

\bibitem{Ingason} Ingason, A. S.; Mockute, A.; Dahlqvist, M. \textit{et. al.}; Magnetic Self-Organized Atomic Laminate from First Principles and Thin Film Synthesis,  \textit{Phys. Rev. Lett.} \textbf{2013} 110, 195502.

\bibitem{Magnuson0} Magnuson, M.; Mattesini, M.; Bugnet, M.; Eklund, P., The origin of anisotropy and high density of states in the electronic structure of Cr$_2$GeC by means of polarized soft X-ray spectroscopy and ab initio calculations,  \textit{J. Phys. Cond. Mat.} \textbf{2015} 27, 415501.

\bibitem{Grossi} Grossi, J.; Shah, S. H.; Artacho E.; Bristowe, P. D., Effect of magnetism and temperature on the stability of (Cr$_{x}$,V$_{1-x}$)$_{2}$AlC phases,  \textit{Phys. Rev. Mat.} \textbf{2018} 2, 123603.

\bibitem{Lue} Lue, C. S. \textit{et. al.}; NMR study of the ternary carbides M$_2$AlC (M=Ti,V,Cr),  \textit{Phys. Rev. B}  \textbf{2006} 73, 035125.

\bibitem{AA} Abdulkadhim, A. \textit{et. al.}; Crystallization kinetics of amorphous Cr$_{2}$AlC thin films,  \textit{Surf. Coat. Technol.} \textbf{2011} 206, 599.

\bibitem{Schneider} Schneider, J. M. \textit{et. al.}; \textit{Ab initio} calculations and experimental determination of the structure of Cr$_{2}$AlC,  \textit{Solid State Commun.} \textbf{2004} 130, 445.

\bibitem{Stohr1} St\"{o}hr, J., Exploring the microscopic origin of magnetic anisotropies with X-ray magnetic circular dichroism (XMCD) spectroscopy,  \textit{J. Magn. Magn. Mat.} \textbf{1999} 200, 470.

\bibitem{Stohr2} J. St\"{o}hr and H. C. Siegmann, \textit{Magnetism}, \textbf{2006} Springer Verlag.

\bibitem{Durr} D\"{u}rr, H. A.; \textit{et al.}; A Closer Look Into Magnetism: Opportunities With Synchrotron Radiation,  \textit{IEEE Transactions on Magnetics} \textbf{2009} 45, 15 (2009). 

\bibitem{Maurizio} Mattesini, M.; Magnuson, M.; Electronic correlation effects in the Cr$_{2}$GeC M$_{n+1}$AX$_{n}$ phase,  \textit{J. Phys. Cond. Mat.} \textbf{2013} 25, 035601.

\bibitem{Ramzan1} Ramzan, M.; S. Lebegue, S.; Ahuja, R., Correlation effects in the electronic and structural properties of Cr$_{2}$AlC,  \textit{Phys. Status Solidi RRL} \textbf{2011} 5, 122.

\bibitem{Ramzan2} Ramzan, M.; Lebegue, S.; Ahuja, R., Electronic and mechanical properties of Cr$_{2}$GeC with hybrid functional and correlation effects,  \textit{Solid State Commun.} \textbf{2012} 152, 1147.

\bibitem{Jaouen} Jaouen, M.; Bugnet, M.; Jaouen, N.; Ohresser, P.; Mauchamp, V.; Cabiocâh, T.; Rogalev, A., Experimental evidence of Cr magnetic moments at low temperature in Cr$_{2}$A (A= Al, Ge) C,  \textit{J. Phys.: Cond. Mat.} \textbf{2014} 26, 176002.

\bibitem{Zhou} Zhou, W.; Liu, L.; Wu, P., First-principles study of structural, thermodynamic, elastic, and magnetic properties of Cr$_{2}$GeC under pressure and temperature,  \textit{J. Appl. Phys.} \textbf{2009} 106, 033501.

\bibitem{Mn-pek} Magnuson, M.; Duda, L.-C.; Butorin, S. M.; Kuiper, P.; Nordgren, J., Large magnetic circular dichroism in resonant inelastic x-ray scattering at the Mn $L$-edge of Mn-Zn ferrite,  \textit{Phys. Rev.} \textbf{2006} 74, 172409.

\bibitem{Eklund2} Eklund, P.; Bugnet, M.; Mauchamp, V.; Dubois, S.; Tromas, C.; Jensen, J.; Piraux, L.; Gence, L.; Jaouen, M.; Cabioh, T., Epitaxial growth and electrical transport properties of Cr$_2$GeC thin films,  \textit{Phys Rev. B} \textbf{2011} 84, 075424.

\bibitem{Kovalik} Kowalik, I. A.; \"{O}hrwall, G.; Jensen, B. N.; Sankari, R.; Wallén, E.; Johansson, U.; Karis, O.;  Arvanitis, D., Description of the new I1011 beamline for magnetic measurements using synchrotron radiation at MAX-lab,  \textit{Journal of Physics: Conference Series} \textbf{2010} 211, 012030.

\bibitem{QE} Quantum-ESPRESSO, see http://www.quantum-espresso.org and http://www.pwscf.org.

\bibitem{Laskowski} Laskowski, R.; Blaha, P., Understanding the $L_{2,3}$ x-ray absorption spectra of early $3d$ transition elements,  \textit{Phys. Rev. B} \textbf{2010} 82, 205104-205110.

\bibitem{Thole} Thole, B. T.; Carra, P.; Sette, F.; van der Laan, G., X-ray circular dichroism as a probe of orbital magnetization,  \textit{Phys. Rev. Lett.} \textbf{1992} 68, 1943. 

\bibitem{Chen} Chen, C. T.  \textit{et al.}; Experimental confirmation of the X-ray magnetic circular dichroism sum rules for iron and cobalt,  \textit{Phys. Rev. Lett.} \textbf{1995} 75, 152.

\bibitem{wien2k} Blaha, P.; Schwarz, K.; Madsen, G. K. H.; Kvasnicka, D.; Luitz, J., \textbf{2001} An Augmented Plane Wave + Local Orbitals Program
for for Calculating Crystal Properties (Karlheinz Schwarz, Techn. Universit\"at Wien, Austria), ISBN 3-9501031-1-2.

\bibitem{dft1} Hohenberg, P.; Kohn, W., Inhomogeneous Electron Gas \textit{Phys. Rev.} \textbf{1964} 136 B864-B871.

\bibitem{dft2} Kohn, W.; Sham, L. J., Self-Consistent Equations Including Exchange and Correlation Effects \textit{Phys. Rev.} \textbf{1965} 140 A1133-A1138.

\bibitem{GGA_WC1} Wu, Z.; Cohen, R., More accurate generalized gradient approximation for solids \textit{Phys. Rev. B} \textbf{2006} 73 235116-235122.

\bibitem{GGA_WC2} Tran, F.; Laskowski, R.; Blaha, P.; Schwarz, K., Performance on molecules, surfaces, and solids of the Wu-Cohen GGA exchange-correlation energy functional \textit{Phys. Rev. B} \textbf{2007} 75, 115131-115145.

\bibitem{Blochl} Bl\"ochl P. E.; Jepsen, O.; Andersen, O. K., Improved tetrahedron method for Brillouin-zone integrations \textit{Phys. Rev. B} \textbf{1994} 49, 16223-16233.

\bibitem{Monkhorst} Monkhorst, H. J.; Pack, J. D., Special points for Brillouin-zone integrations \textit{Phys. Rev. B} \textbf{1976} 13, 5188-5192.

\bibitem{Anisimov1991} Anisimov, V. I.; Gunnarsson, O., Density-functional calculation of effective Coulomb interactions in metals \textit{Phys. Rev. B} \textbf{1991} 45, 7570-7574.

\bibitem{Singh1994} Singh, D., \textbf{1994}, Plane waves, pseudopotentials and the LAPW method, Kluwer Academic, Boston.

\bibitem{Hubbard} Hubbard, J., \textbf{1963} Electron correlations in narrow energy bands \textit{Proc. R. Soc. London} Ser. A 276 238.

\bibitem{B3LYP} Stephens, P. J.; Devlin, F. J.; Chabalowski, C. F.; Frisch, M. J., Ab initio calculation of vibrational absorption and circular dichroism spectra using density functional force fields, \textit{J. Phys. Chem} \textbf{1994} 98, 11623.

\bibitem{MT} Tyson, T. A.; Hodgson, K. O.; Natoli, C. R.; Benfatto, M.; General multiple-scattering scheme for the computation and interpretation of x-ray-absorption fine structure in atomic clusters with applications to SF$_{6}$, GeCl$_{4}$, and Br$_{2}$ molecules, \textit{Phys. Rev. B} \textbf{1992} 46, 5997.

\bibitem{FDMNES} Bunau, O.; Joly Y., Self-consistent aspects of x-ray absorption calculations, \textit{J. Phys. Cond. Mat.} \textbf{2009} 21, 345501. 

\bibitem{Pardini} Pardini L.; Bellini V.; Manghi F.; Ambrosch-Draxl, C., First-principles calculation of X-ray dichroic spectra within the full-potential linearized augmented plane wave method: An implementation into the Wien2k code, \textit{Comput. Physics Comm.} \textbf{2012} 183, 628.

\bibitem{NA} Abdurakhmanova, N. \textit{et al.}; Superexchange-mediated ferromagnetic coupling in two-dimensional Ni-TCNQ networks on metal surfaces,  \textit{Phys. Rev. Lett.} \textbf{2013} 110, 027202.


\bibitem{Tomaz} Tomaz, M. A.; Antel, W. J.; O'Brien, Jr. W. L.; Harp G. R.; Orientation dependence of interlayer coupling and interlayer moments in Fe/Cr multilayers,  \textit{Phys. Rev. B} \textbf{1997} 55, 3716.

\bibitem{Alayo} Alayo, W.; Tafur, M.; Xing, Y. T.; Baggio-Saitovich, E.; Nascimento, V. P.;  and Alvarenga, A. D., Study of the interfacial regions in Fe/Cr multilayers,  \textit{J. Appl. Phys.} \textbf{2007} 102, 073902. 

\bibitem{Schneider,} Schneider, J. M.; Sun, Z. M.; Mertens, R.; Uestel, F.; Ahuja, R., Ab initio calculations and experimental determination of the structure of Cr$_{2}$AlC,  \textit{Solid State Commun.} \textbf{2004} 130, 445.

\bibitem{Luo} Luo, W.; Ahuja, R., Magnetic Fe$_{n+ 1}$AC$_{n}$ (n= 1, 2, 3, and A= Al, Si, Ge) phases: from ab initio theory,  \textit{J. Phys. Cond. Mat.} \textbf{2008} 20, 064217.

\bibitem{Heyd} Heyd, J.; Scuseria, G. E.; Ernzerhof, M., Hybrid functionals based on a screened Coulomb potential,  \textit{J. Chem. Phys.} \textbf{2003} 118, 8207.

\bibitem{Sun} Sun, W.; Luo, W.; Ahuja, R., Role of correlation and relativistic effects in MAX phases,  \textit{J. Mater. Sci.} \textbf{2012} 47, 7615.

\bibitem{Kim} Kim, D. J., The electron-phonon interaction and itinerant electron magnetism,  \textit{Phys. Rep.} \textbf{1988} 171, 129.

\bibitem{Liu} Liu, Z.; Waki, T.; Tabata, Y.; Nakamura, H., Mn-doping-induced itinerant-electron ferromagnetism in Cr$_{2}$GeC,  \textit{Phys. Rev. B} \textbf{2014} 89, 054435.

\bibitem{Sole} Del Sole, R., Girlanda, R.; Optical properties of semiconductors within the independent-quasiparticle approximation, \textit{Phys. Rev. B} \textbf{48} 11789.

\bibitem{Scherz} Scherz, A.; Wende, H.; Sorg, C.; Baberschke, K.; Minr, J.; Benea, D.; H. Ebert, H., Limitations of integral XMCD sum-rules for the early $3d$ elements,  \textit{Physica Scripta} \textbf{2005} T115, 586.

\end{thebibliography}
\end{document}